# A smart building semantic platform to enable data re-use in energy analytics applications: the Data Clearing House.

Daniel Hugo, John McCulloch, Akram Hameed, Will Borghei,
Martin Grimeland, Verity Felstead, Mark Goldsworthy

Commonwealth Scientific and Industrial Research Organisation (CSIRO); firstname.lastname@csiro.au

## ABSTRACT

*Systems in the built environment continuously emit time series data about resource usage (e.g., energy and water), embedded electrical generation/storage, status of equipment, patterns of building occupancy, and readings from IoT sensors. This presents opportunities for new analytics and supervisory control applications that help reduce greenhouse gas emissions due to energy demand, if the barrier of data heterogeneity can be overcome. Semantic models of buildings – representing structure, integrated equipment, and the many internal connections – can help achieve interoperable data re-use by describing overall context, in addition to metadata. In this paper, we describe the Data Clearing House (DCH), a semantic building platform that hosts sensor data, building models, and analytics applications. This fulfills the key phases in the lifecycle of semantic building data, which includes: cost-effective ingestion of Building Management System (BMS), IoT, metering and meteorological time series data from a wide range of open and proprietary systems; importing and validating semantic models of sites and buildings using the Brick Schema; interacting with a discovery API via a high-level domain-specific query language; and deploying applications to modelled buildings. Having onboarded multiple buildings belonging to our own organisation and external partners, we are able to comment on the challenges to success of this approach. As an example use-case of the semantic building platform, we describe a measurement and verification (M&V) application implementing the 'whole facility' (Option C) method of the International Performance Measurement and Verification Protocol (IPMVP) for evaluating electrical metering data. This compares energy consumption between nominated baseline and analysis time periods, to quantify the energy savings achieved after implementing an intervention on a site. For each target building, the platform permits dynamic discovery of relevant time-series sensor data, enabling a more automated application deployment process.*

## CONTEXT OF THE DATA CLEARING HOUSE (DCH) PLATFORM

The Data Clearing House project began as a collaboration between CSIRO business units (Energy, Data61), the Australian Institute of Refrigeration, Air Conditioning and Heating (AIRAH), and a range of external partner organisations, with a vision to enable new applications that reduce energy demand in the buildings sector, by breaking down conventional information silos that hinder interoperability and putting building owners and managers back in control of gaining value from data generated by their assets. The value propositions of the DCH are a) sensor data becomes more useful when consolidated; b) semantic models of buildings can serve many applications; and c) applications are more powerful when deployment effort is minimised.

### Sensor data management

Time series data generated by buildings includes quantitative scalar data such as sensor observations and control setpoints, and qualitative data such as statuses, alarms, and commands. This data is often stored across multiple systems, with varying degrees of accessibility. This means that the first task for the DCH is to act as a unified repository of time series data, with a simple API for exploitation by algorithms.

### Semantic building models

Semantic models of buildings are the DCH's mechanism for representing data in context. While basic metadata (e.g., observed quantity, unit of measurement) are useful, semantic models go even further by describing the holistic *context* in which sensor data exists. This includes the locations that exist within a building's structure, the installed equipment that provides services, and the paths through which air, water, and energy are conveyed.

*Applications*

'Applications' in the DCH are software packages that provide value to users through the algorithms they implement. Separating app development from app deployment means that developers should not need to know specific target sites ahead of time. Apps query building models to discover relationships and data as required, including determination of when a given building model does not contain required information. The purpose of the DCH is to provide this mechanism of re-deployable applications.

## OVERVIEW OF THE DCH PLATFORM ARCHITECTURE

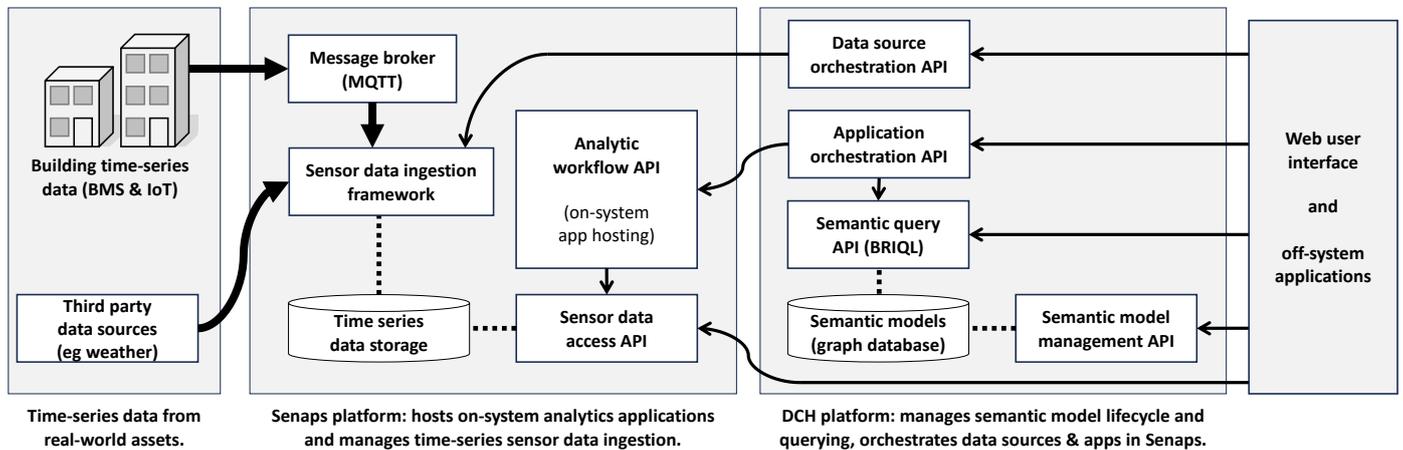

**Figure 1: overview of the DCH platform architecture**

Figure 1 shows the high-level components that make up the Data Clearing House platform. Some of the functionality offered by the DCH depends on an adjacent service, the 'Senaps' sensor data analytics platform (Coombe et al. 2017). Senaps is a mature platform, developed by CSIRO, and commercialised by Eratos Pty Ltd. Its features of relevance to the DCH include:

- Data storage, including time series scalar data with basic metadata (quantity kind, unit of measure, etc).
- A framework for managing data ingestion, via configurable 'data sources'.
- An analytic workflow service, for hosting and running arbitrary analytic software.
- An API to access sensor data (used by the workflow service and exposed externally).
- Role-based access control over users and groups, and an OAuth2 identity provider.

'Workflows' in Senaps are a versatile mechanism for running analytic software as needed in isolated containers, with access to time series sensor data secured by Senaps' role-based access control. Workflows are composed of 'operators' (reusable software components) and may emit new streams back into the time series database.

The DCH platform, through its APIs, manages the lifecycle of semantic models and provides a semantic query service. It also interacts with Senaps to orchestrate data sources, manage the lifecycle of on-system hosted applications, and provides role-based authentication and authorization services for semantic models and data.

## BMS AND IOT INGESTION IN DCH

The DCH aims to be neutral towards data formats, hardware, and external software, and can communicate with the outside world through most popular protocols. Preferred interfaces are MQTT and AMQP, but REST API and FTP options are also available. These interfaces allow the DCH to ingest both metadata and time series data from a wide range of third-party cloud services or on-site hardware. Some options are fully self-serviced by end users, while other options require the DCH team to assist commissioning.

### Low-level on-site communication

The assets typically found in any modern commercial building can be grouped by their communication mechanisms. We distinguish between Building Management System (BMS) assets, Legacy assets, and Internet-of-Things (IoT) assets.

BMS assets are the equipment used in a building's essential operation, e.g., chillers, boilers, water pumps, lighting services and Heating, Ventilation, and Air Conditioning (HVAC) equipment etc. Typically, these devices communicate using the BACnet protocol. BMS systems often have the capability to forward the data to the internet via one of the supported protocols, however sometimes additional gateway devices must be installed on the network to capture and forward data to the DCH.

Where building assets are not BACnet compatible, protocols such as Modbus are common. Modbus does not support a discovery mechanism, so onboarding Modbus devices requires the system integrator to configure a gateway to poll specific devices and forward information to the DCH.

Finally, IoT sensors and other equipment, where vendors provide APIs to access equipment data, have become popular assets both in residential and commercial buildings. These devices can be anything from Temperature/Humidity/Luminosity/Motion (THLM) sensors to smart utility meters or inverter systems. In general, these devices transfer data to some endpoint accessible to device owners via the internet.

### Gateway to the DCH

As discussed above, the edge device or gateway must have a connection to the internet for data uplink. Most industry gateways support MQTT. If MQTT is not available, devices may need to interface via the Senaps REST API or FTP. A typical configuration used with the DCH is a Tridium JACE-8000 device (the gateway in this case) installed on a site with network access to the BMS, utility meters, and the internet. Data is polled at regular intervals on both BACnet and Modbus and then sent to the DCH using MQTT.

### Third-Party platform bridge

If data is already onboarded to a third-party cloud platform, the DCH can use various protocols to access and ingest that data. This is the typical IoT solution as vendors typically provide data via some internet accessible server. Depending on the capabilities of the third-party platform, data can be ingested into the DCH using MQTT, Senaps REST API or FTP. These are also the mechanisms that can be used to get data into the DCH via some third-party (non-vendor) platform. Operating examples include:
- subscription to the TTS (The Things Stack) MQTT broker to ingest data from LoRaWAN IoT sensors
- ingestors where data is pushed in from other third-party systems using the Senaps REST API
- ingestors where FTP is used to capture CSV data in NEM12 format, the Australian Energy Market Operator standard for metering data interchange (AEMO, 2022).

### The DCH payload formats

Although the DCH can be extended to accept any payload format, a JSON schema has been released to help clients standardise their data payloads (https://bitbucket.csiro.au/projects/SBDCH/repos/bms-json/browse). Operational examples exist of gateways that send data to the DCH in both the DCH's standardised format and more bespoke formats. As new formats are developed, they can be shared between clients. For example, a new data parser was added to support OpenADR payloads (IEC, 2018) via MQTT. The use of open-source protocols and payloads is encouraged by the DCH.

### Security

The DCH encourages all users, regardless of the nature of their buildings, to use encryption for all communication to and from the DCH. This is supported via all ingestion mechanisms.

## THE DATA CLEARING HOUSE'S BUILDING SERVICE LAYER (BSL) API

The DCH Platform's semantic model functionality is provided by the Building Service Layer (BSL). This exposes a RESTful web API with endpoints for managing model lifecycle and querying. The BSL also exposes endpoints for managing application lifecycle.

### *Meta-objects*

Multiple agencies or businesses may own data in the DCH platform. To orchestrate multi-tenant access control, the uppermost level of segregation is an 'Organisation'. Users with administration rights for an Organisation can manage access rights for their personnel. The next two levels of segregation are 'Site' and 'Building' meta objects, instances of which can be associated with basic metadata (geolocation, address, and cadastral information), semantic models, time series data, and deployed applications. 'Sites' represent land parcels, external spaces, and facilities or systems that serve Buildings. 'Buildings' represent the permanent enclosed structures constructed on Sites.

*Models augment metadata with context*

Semantic models in the DCH describe the composition of Sites and Buildings, using concepts from the Brick Schema (Balaji et al, 2016). Brick classifies a hierarchical vocabulary of types, including physical locations (e.g., wings, floors, rooms), logical locations (zones), physical equipment, and systems (ensembles of related equipment). Every stream of time series data ingested into the DCH can be associated with an instance of a Brick 'Point' class (representing a sensor, status, setpoint, parameter, command, or alarm information) in a building model. The entities of a semantic model (instances of locations, equipment, or points) can be annotated with 'entity properties' to describe their characteristics.

Entities in models are connected via Brick's relationships. These exist as reciprocal pairs and describe composition of entities (`hasPart`/`isPartOf`), their position (`hasLocation`/`isLocationOf`), transfer of matter or energy (`feeds`/`isFedBy`) and association of point to location or equipment (`hasPoint`/`isPointOf`). Relationships are transitive; this means, for example, if a wing has several floors, and those floors each have many rooms, then the rooms belonging to the wing can be found by transitively following '`hasPart`' from wing to room via floor. Building models are thus graphs of nodes connected via directed heterogenous edges. In the case of 'feeds', sequence of edges may be cyclic. A very simple example building model, showing HVAC components, is given in Figure 2.

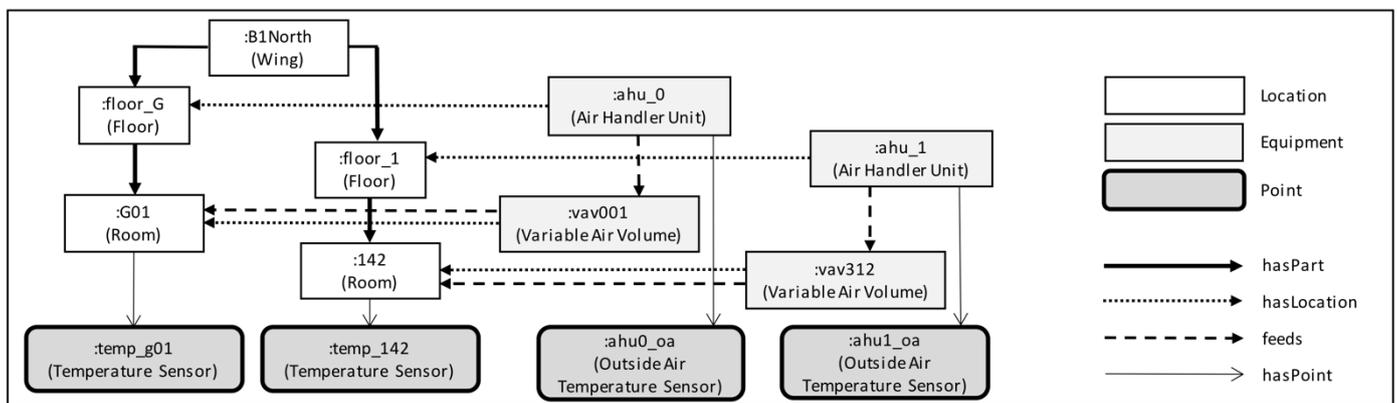

**Figure 2 Modelling an HVAC system using Brick**

*Semantic querying*

The purpose for storing graph-based semantic models of sites and buildings is to permit discovery of entities, especially 'points', that match an arbitrary topology and/or other criteria. Several query languages exist for querying graph-like data, including Cypher (Francis et al, 2018), Gremlin (Rodriguez, 2015), and SPARQL (W3C, 2013), each of which has its own intended use-cases and limitations. Since the emerging semantic buildings community has generally favoured RDF-based schemas and tooling, it made sense to use a SPARQL-based database with the DCH. We chose Blazegraph, but any RDF triplestore supporting named graphs would suffice.

*Why not just use SPARQL?*

SPARQL is an expressive graph query language for RDF graphs. However, some of its traits are poorly suited to direct use in a web API for third parties, either because of security or because of developer ergonomics. The most significant are:
- Access Control: SPARQL-based databases generally do not restrict read access at the granularity of named graphs, meaning some kind of external constraint is required.
- Quality of Service: In a cloud platform, mitigation of inadvertently inefficient queries is important.
- Target Audience Useability: Application developers in this domain, and in Web systems generally, anticipate using modern object-oriented APIs, while SPARQL's similarity to SQL adds potential friction.
- Flexibility: SPARQL queries must be provided verbatim every time they are invoked; they cannot be stored and invoked like functions with arguments.

## BRIQL, a query API for the Brick Schema

We observed that SPARQL queries on models using Brick schema typically follow a common pattern, starting by matching a set of building entities by type and/or topology, and finishing by extracting the attributes of those entities. To relieve developers of this repetitive workload, we designed a query API, named BRIQL, with features to automate these steps and expose query solutions in object-oriented representation. Its key features are:

- Each BRIQL 'variable' represents a building entity (location, equipment, zone, or point) to be matched in query solutions, rather than a *single attribute of an entity* as with SPARQL variables.
- For each BRIQL variable, metadata attributes can optionally be requested. The most useful of these is 'pointinfo', a shorthand to request "for this entity, also find any attached Points and their metadata".
- To constrain a variable, its type can be specified (a) by exact reference to a Brick class, (b) by choosing a parent class in Brick's hierarchy, (c) indirectly by a list of relevant tags with which Brick's classes are each associated, or (d) indirectly by properties of the instance. Topological constraints expressing inter-node connectivity are also available.
- BRIQL queries can be stored on the DCH platform, for later invocation by reference.
- Like SPARQL, a 'describe' mode acts as a shorthand to get all information about an individual entity.

## Lifecycle of an invocation in the BRIQL API

A BRIQL invocation specifies a BRIQL query (either literally with the invocation, or by reference), a list of sites and/or building models to examine, and optional arguments (bindings to override input variables' default values). After the BSL checks access for requested models, the invocation is dynamically translated to a SPARQL query matching all requested topologies and types and binding requested metadata. In this translation, SPARQL graph pattern sequence is heuristically optimised for time and memory impact. After the graph database executes the SPARQL query, the raw solution is parsed into a list of entities (each with any requested metadata), and a table of solutions where each column of each row contains a reference to an item in the list of entities. The entity list and solution table are composed into a BRIQL response and returned in JSON format to the invoking application.

## BRIQL query example

Consider an HVAC validation application that needs to calculate a differential between the outside air intake temperature sensors of air handler units (AHUs) and the ambient temperature sensor of each room supplied (via whatever route) by those AHUs. To discover these associated pairs, a developer writes a BRIQL query such as this:

```
  "variables": [
    { "name":"ahu",
      "output": true,
      "brick_type": {"match":"isa", "type":"AHU"},
      "fetch": ["id","pointinfo"],
      "fetch_points": [{"match":"tags", "tags": ["Outside", "Temperature", "Sensor"]}]
    },
    { "name":"room",
      "output":true,
      "brick_type": {"match":"isa", "type":"Room"}
      "fetch": ["id","pointinfo"],
      "fetch_points": [{"match":"tags", "tags": ["Temperature", "Sensor"]}]
    }
  ],
  "query": { "paths":[
    {"from_ref":"ahu", "properties":[{"property":"feeds", "min":1}], "to_ref":"room"}
  ]}
```

This defines two variables named 'ahu' and 'room'. The variable declarations each specify a brick type to match, request that information about attached points be fetched, and constrain point types using Brick's class tags. Finally, the query requires that the 'ahu' variable be transitively linked (through at least one 'feeds' connection) to the 'room' variable in every solution. When invoked upon the example building model of Figure 2, this would discover the associations between sensors 'ahu0_oa' and 'temp_g01', and between 'ahu1_oa' and 'temp_142'. While queries of far more complicated topology are possible, this example illustrates the concise yet expressive nature of BRIQL.

## APPLICATIONS: BRINGING ALGORITHMS TO THE DATA

The DCH platform helps building owners get value from their data by offering both built-in and third-party applications. To make this model viable, the DCH offers the building application developer community a set of tools to bring their analytic algorithms to market. Developers use the DCH SDK to develop software that can analyse and enrich a facility's data. Using the BRIQL API, developers can describe the data to be accessed from the facility and how to connect the data to the software. Data produced by the software can be presented through customised Grafana dashboards, where the application developer can present analysis outcomes using a range of graphical widgets.

### Application packaging

A DCH application is built from a set of operators (reusable components for processing sensor data), a declaration of the application context, instructions on the flow of data, a set of queries to run against sites and buildings, and a blueprint for how to render the application frontend. The workflow framework allows operators to be written in Python, R, or MATLAB. Through the application context declaration, the application developer controls the version of the application, provides name, description, and can specify what memory and CPU resources it requires. The application is deployed to the DCH's 'application store' by creating an archive of all the components of the application, including the operators, and uploading through the DCH API. Once applications have been uploaded, they are ready for users to use (or install) on their sites or buildings.

### Application lifecycle

On-platform DCH applications are run as 'workflows' in the Senaps Analysis Service. When a user requests that an application be deployed for a set of sites/buildings, the DCH platform uses the application's discovery query to determine the time-series data streams to use and instructs Senaps to create a new workflow instance. A Docker container is instantiated in the virtual private cloud, to actually execute the application's code. The container context allows access to the discovered time-series data but prevents any external communication. This solves a key trust concern, since DCH users do not need to worry about the possibility of malicious apps exfiltrating sensitive data.

### Off-platform applications

Because models and the query engine can be accessed via API, off-platform applications can also be constructed using any tooling capable of interacting with those APIs. Off-platform applications have the potential to provide a rich user interface or experience. The trade-off for this versatility is accepting that, unlike on-platform applications, data will leave the DCH platform.

## WEB INTERFACE

The DCH includes a web front-end with which users can onboard their buildings, models and data, without a technical background or in-depth understanding of the Brick schema. The user interface uses the DCH and Senaps APIs to allow building owners and managers to connect sensor data, view time series data charts, conduct semantic model management install applications on sites or buildings and view application results. Authorised users can also manage role-based access control for their colleagues at a site or building granularity.

One of the platform's key features is a model editor, which enables users to create and manage semantic models of their buildings. Through this, offline models can be uploaded and validated, and draft models can be explored and edited. Users can 'publish' models to mark them as ready for use. The editor makes it possible for users to interact with the lifecycle of Brick models without ever needing to edit raw RDF files.

Semantic building models can be large, with tens of thousands of entities. A graph-based visualiser provides an interactive network graph for exploring and searching the model data. This model visualiser is built with AntV's G6 graph engine, utilising a force-directed layout which has been extensively customised for the DCH's considerations. It is aware of the Brick ontology and takes advantage of visual properties such as colour and iconography to assist users in quickly understanding information and recognising patterns. It permits progressive dynamic discovery of entities to reduce the performance concerns that arise when interrogating large models.

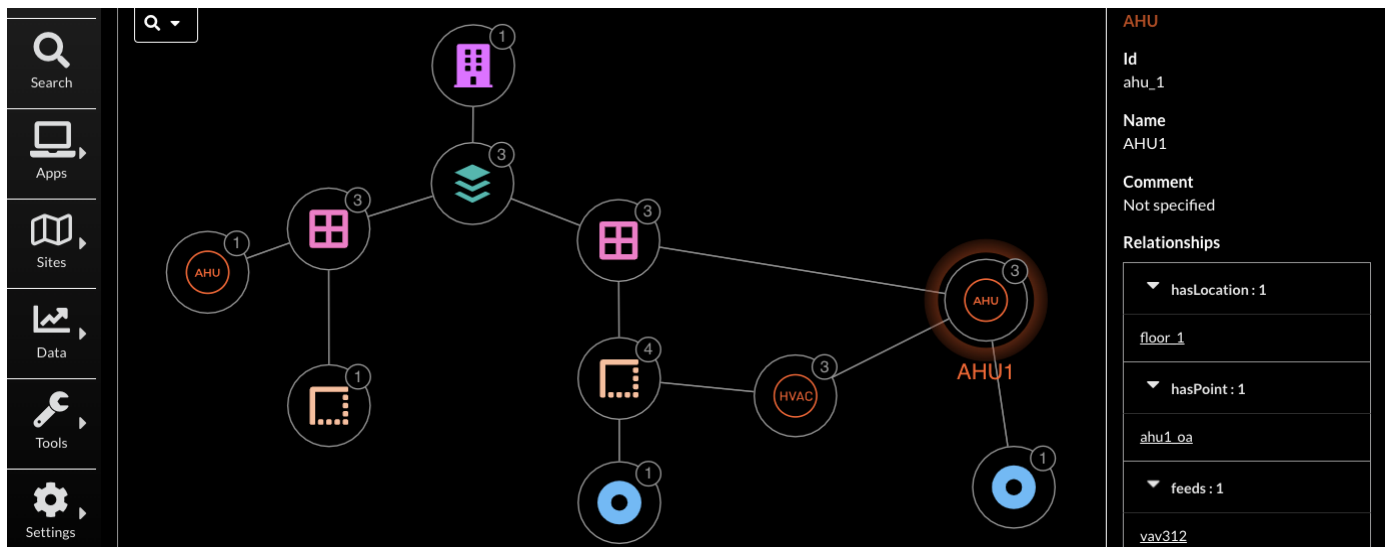

**Figure 3: the DCH model visualiser, showing partial expansion of the building model given in Fig. 2.**

## DCH USE CASE: A MEASUREMENT AND VERIFICATION APPLICATION

As our world attempts to achieve net zero $CO_2$ emissions and avert climate change, consistent mechanisms for validating energy-saving interventions will become increasingly important to policy implementation. In the built environment sector, standardised Measurement and Verification (M&V) algorithms exist to determine actual savings of energy, cost and/or atmospheric impact that are attributable to some intervention or change to building equipment and/or operation.

The International Building Measurement and Verification Protocol (IBMVP) defines four M&V methodologies that are possible depending on the specific case, available data, time, and budget (EVO, 2022). One of these, the 'whole facility' method known as Option C, is becoming increasingly attractive since high-resolution (i.e., sub-hourly) metering data is now available in many buildings. Advanced analytics makes deployment easier and enables identification of even smaller energy savings over shorter time-frames than was previously possible. This is referred to as M&V 2.0 and is an active area of research (EVO, 2019).

IBMVP M&V Option C uses data over a specified period (known as the baseline period) and compares the predictions of this model with measurements taken over a second period (known as the analysis period). The baseline and analysis periods are typically chosen to be representative of operations before and after some alteration to the site.

Example use cases for Option C can include:
- Verification that equipment upgrades or control system changes had the desired outcomes.
- Monthly-to-month comparison of site energy use.
- Long term site energy use trend analysis.

The key differences between an M&V-based process and a simple before/after energy use comparison are:
- Normalising for changes in the weather and shutdown or holiday periods,
- Accounting for temporal gaps in the data,
- Adjusting for non-routine changes to site operation, and
- Providing confidence bounds on estimated energy savings.

The core of the method is based on a decomposition of the site's overall energy consumption into baseload and temperature dependent components. Overall energy consumption in the M&V context means total energy used in the building, which is independent of any onsite generation. Traditional implementation of the algorithm on a site requires manual analysis of installed metering and available metering data for the site to find or calculate that overall energy consumption. Complexities such as data availability and arrangement of generation metering are crucial factors in that decision making. That energy data is then fed into the Option C algorithm with other information such as ambient temperature and occupancy records to determine the result. With the DCH, where the objective of the platform is for applications like M&V to be applied to sites *without* prior analysis of the site (such as manual meter hierarchy discovery), the M&V application needs to algorithmically discover that information prior to running the core analytics. This discovery is done by leveraging the DCH's BRIQL API applied to the semantic model of the site and buildings.

Adding this query component to an application is the necessary cost of transitioning applications from the traditional processes, where application developers are involved in commissioning the application on each building, to a platform-based approach where the application is installed or commissioned on sites without any per-site developer overheads. Because physical systems, metering arrangements and data availability vary widely from site to site, determining the correct queries and logic to determine the correct set of data to use for M&V is not trivial. We argue that the cost-benefit of this transition will be positive for application developers as scale (meaning number of onboarded buildings / potential clients) increases in these semantic based platforms.

*Metering discovery query*

To find the correct set of data, we must find both the correct set of meters as well as the correct metering data from those meters. Finding the correct meters is about examining the metering hierarchy to determine which (set of) meters will provide coverage of all energy consumption on the site, as well as the subsequent set of downstream metering that can also identify generation that feeds into the site or out to the grid. Finding the correct metering data involves selecting meter points for available energy data or power data, discriminating between three-phase data or single-phase data where both are present, identifying data with the correct complexity (real, reactive, apparent) and arithmetic sense (export, import, net). A final complication is the need to assess the completeness of data available to run the app. While this sounds straightforward, the National Australian Built Environment Rating System (NABERS, a "simple, reliable, and comparable sustainability measurement rating system for buildings"), has a 100-page document specifying the rules to be applied in various building and metering scenarios to ensure the system is fit for purpose (NABERS, 2023).

Figure 4 shows a partial example metering hierarchy for onboarded metering for a site modelled in the DCH. While this model is not complete (e.g., most assocaited locations, equipment and points are not shown), it is typical of real-world semantic models. Domain experts can examine such hierarchies, explore data availability, and quickly identify the set of consumption and generation points, and how to combine them, to provide overall energy consumption to the core M&V algorithm. In Figure 4, site consumption can be calculated by summing data from meters 'Supply 1' and 'Supply 2', accounting for generation from 'B060G' (but not 'B501G'). However, if one is only interested in Building 501 consumption then the correct method is to add 'Supply1' and 'Supply 2' and subtract 'B061' and 'B062' data. This tacit knowledge and understanding of the Brick Schema and semantic models is essential to developing the heuristics for a M&V application, that subsequently get translated by the app developer into a combination of BRIQL queries and logic, resulting in an application that uses one set of 'rules' to successfully discover the correct set of data to use when the application is installed on any building model.

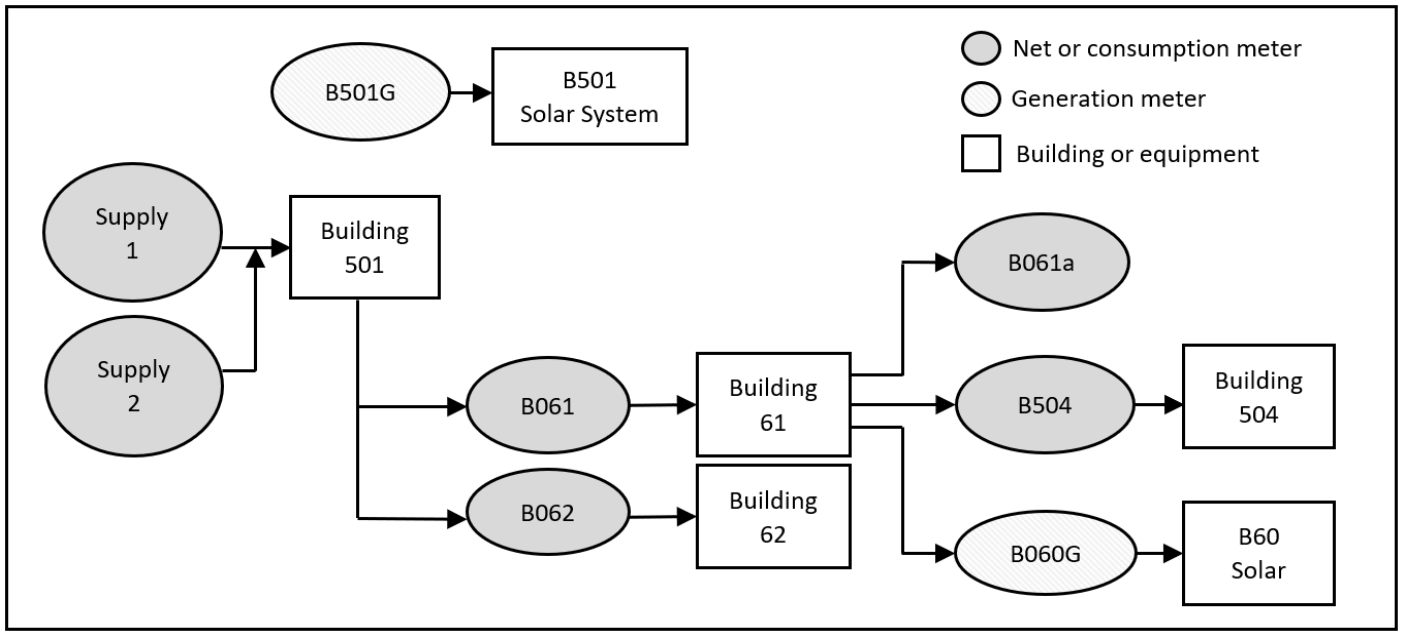

**Figure 4: Example meter hierarchy**

## DISCUSSION: OBSERVATIONS AND CHALLENGES

This paper has presented an overview of the DCH platform for deploying applications to exploit buildings sector data through semantic modelling, motivated by energy reduction use cases. The production DCH system is in use by our own organisation's facilities management business unit, and by a growing cohort of external partner organisations. At the time of writing, the number of 'onboarded' buildings (with both live time-series data and semantic models) has passed 50. In addition to platform development, we have ourselves undertaken roles including data source configuration, semantic modelling, and application development. From this perspective, we here describe some of the challenges that were encountered, some of which are the focus of ongoing research efforts.

### Query API evolution

While BRIQL is a useful first step towards interrogating semantic building models without the overheads of SPARQL, we recognise that further work is required. Although BRIQL's concise idioms and object-oriented response are productivity gains over raw SPARQL, its JSON serialisation is not ideal for a query language. An area of ongoing research is a graphical query composer for semantic models in the buildings sector. This is more complicated than simply describing a topology of nodes, as queries must allow negation and ordering of constraints. The standard we still aim to achieve is that any application developer (or even non-expert data owner) should be able to compose and understand semantic queries with a minimal learning curve.

### Data Health

The DCH aims to be a platform to enable the deployment of analytics applications where the application author need not have conducted an evaluation on the quality of data owned by a particular site or building prior to deploying their application. This assumption generates serious challenges for data providers like BMS contractors and data managers or maintainers, including the DCH platform.

In the DCH ecosystem, we describe *Data Health* in the context of traditional definitions of *Data Quality*, such as the following: 'data that are fit for use by data consumers' (Wang et al. 1996). Use of this alternative term has been adopted with an aspiration to provide a *data health metric* that may include components that are not traditionally the purview of *data quality*, such as factors external to the dataset (e.g. equipment failures in remote publishers). That traditional definition also highlights the perennial challenge of different applications having different requirements on data quality, meaning 'fit for purpose' is somewhat subjective.

There are many roles responsible for ensuring data and metadata in the DCH is both consistent and valuable to application developers. Data health, then, is not just the domain of data providers, but also the domain of the platform itself, as well as the domain of model maintainers. Data health challenges specific to the data provider include the following non-exhaustive list, each of which has been encountered during the prototype phase of the DCH platform.

### *Equipment faults causing data outages*

Faults in building equipment occur infrequently, but often enough that they present a challenge for data providers and data maintainers to deal with. One might assume that if a piece of equipment fails to operate altogether, then no data will be received by the DCH. This has been shown to *not* be the case. Many edge devices operating as building gateways will report sensor readings even when a field device has not reported within an appreciable window of time. The readings reported could then be classified in several ways. Do they represent a problem of accuracy since they do not represent the true state of the equipment? Do they represent a problem of validity, or completeness, in that they may represent the state of the equipment prior to the fault occurring (stale data), or is there a data integrity problem in that the true state of the equipment is being masked by the reporting equipment? Each of these questions is raised when the DCH team encounters failure conditions of remote equipment.

### *Equipment behaviour and its impact on data quality*

The DCH sets as a goal to allow ingestion of data from any device or service that can communicate with supported open protocols. This goal, while noble, leads to a series of data quality issues that sometimes cannot be anticipated. In one example, faulty software on an edge device resulted in the clock on the device drifting over time. This resulted in incorrectly timestamped time series data being sent to the DCH. Since the platform has no knowledge of when the data was sampled, it can only trust the data as reported by the edge device. This presents a problem when a clock has drifted significantly, as users who rely on the data to be accurate have no way of knowing that a clock has drifted (except in a simple case where the values are reported to be in the future). Control applications are at risk of making decisions based on untimely data, potentially leading to invalid decisions, or inappropriate control signal recommendations.

An additional concern with equipment is when a fault results in data that is out of expected range for a given parameter. This is a more insidious problem than the simple case where a sensor provides no value, as the consumers of data may not be aware that the equipment is faulty and that its readings cannot be trusted. In cases like that of a thermocouple, a simple data checking approach can determine that a value of 0 degrees Celsius is not expected in an office building. However, if the device was installed in a freezer, this could be a legitimate value for the equipment.

In many cases, resolution of issues with on-site equipment requires a contractor to attend and perform a fault diagnosis activity. For the DCH, providing users with the ability to discover and triage issues with their data health is an ongoing concern.

### *Equipment mapping discrepancies*

This third contributor to data health issues in the DCH is one that the platform itself may be able to alleviate in the future. When equipment is commissioned on a site, its control and sensor 'points' are connected to edge devices capable of reading and writing data. These points are then configured in the BMS to give them meaning. A system integrator will take the generic names for edge device points like 'analogue input 1' and map it to a name that is convenient for future use, such as 'outside_air_temperature_sensor'. This mapping activity is largely a manual process, with all the perils that entails. System integrators, even those with significant experience, may still make errors in this manual process, meaning that the name of a point that the BMS reports may not reflect reality.

A variation on this problem arises when building works such as extensions or refits occur, and equipment or sensors are removed or updated. The field units connected to sensors may have their inputs or outputs re-assigned to new equipment. If the integration process is not completed appropriately in the BMS, the point mapping becomes stale or invalid.

The inherent uncertainty that results from a point being mapped incorrectly is passed on to the DCH as part of the 'point classification' process, where a qualified individual links time series data to semantic models. Methods for dealing with this are beyond the scope of this paper but are of great interest to the DCH team.

**Model creation, validation, and maintenance**

Semantic models are part of the DCH data ecosystem, making fit-for-purpose semantic models of buildings an essential requirement. Although model generation processes are out of scope for this paper, it is important to recognise challenges with creation, validation, and upkeep of semantic models.

Creation of models is largely an overhead to building owners (value is derived from *use* of models, not their mere *existence*) so it is important to constrain model scope to match desired applications. Measures for completeness and accuracy of models are difficult to define. To highlight the issue, consider Figure 4. Is that model complete? Is it accurate? Examination identifies that there is no grid-side connection to generation meter 'B501G' shown. Is that because that meter has its own grid connection or is it an omission in that diagram/model and it really connects downstream of meters 'Supply 1' & 'Supply 2'? Automatic determination of such potential omissions is not possible for the general case. Rules around how things should be modelled can help but cannot solve all cases. One solution to the 'completeness' question is the concept of 'semantic sufficiency', whereby application requirements are collated and used to test a model's suitability for certain uses (Fierro et al, 2022). We see this as a useful paradigm for the DCH to adopt in model and application authoring processes.

Ongoing maintenance of models is also a challenge for building owners. Processes must be put in place to ensure that semantic models keep pace with changes to energy, HVAC and other building assets. As touched on below incentivising actors in the ecosystem to do this will be important.

**Providing an incentive**

As previously discussed in this paper, one of the goals of the DCH is to provide a data platform where building managers and operators are in control of their own data and may integrate data from many different providers. We describe here several challenges to data integration relating to non-technical or business matters.

The traditional approach to building data management is that the organisations who manage your BMS also manage your building data on your behalf. This presents a potential barrier to adoption of the DCH in industry. BMS operators are unlikely to release data to the DCH for purely altruistic reasons, as the data in their charge has value to them. Providing incentives to system integrators to participate in the DCH ecosystem is becoming an area of interest to the DCH team.

One challenge to participation is that of protocol use. As discussed earlier, the DCH is made accessible to system integrators by using open protocols. While significant numbers of buildings can provide their data using these sorts of protocols, there remain an equally significant number of buildings that cannot have their data exported to the DCH via the recommended 'open' channels. These 'closed' protocol buildings exchange data internally using proprietary protocols that cannot be integrated with other vendors' equipment without translation devices. For building owners and managers to conenect their data to the DCH, more needs to be done to encourage vendors to provide open access to a customer's own data.

**Cyber Security Considerations**

The DCH has been built using industry standard cyber security approaches wherever possible. Despite this, there are several challenges that arise when users engage with the platform to store their building data as part of typical onboarding scenario.

One learning of the DCH team has been that the technology used in the DCH may not be well understood by security or standard information technology (IT) staff of data providing organisations. Developing an appropriate level of trust with IT services to permit in-house data to be sent to the cloud is crucial to the DCH's success.

In some cases, the cyber security policies of organisations do not allow equipment or software to be integrated into their building control networks, preventing the DCH (or other systems) from being implemented on a particular building. It is also anticipated that users will seek applications that can perform supervisory control in their buildings. As this happens, security and reliability will become much more important than in simpler analytics use cases where no control is possible.

## CONCLUSION

The built environment sector, as a major energy consumer, represents an opportunity to reduce global greenhouse gas emissions, and thereby reduce the impact of climate change. Doing so will require exploitation of sensor and control data that is often siloed inside building management systems, but the heterogeneity of data and a lack of standardised metadata are ongoing barriers to the uptake of new analytic and control applications.

We have described the Data Clearing House (DCH), a platform that addresses these challenges through semantic building models. The DCH is intended as a vendor-neutral solution to the problem of building data integration for analytics at scale. It permits the aggregation and storage of time series data, the creation of semantic models of buildings that reference that time series data, and a novel domain specific query API that facilitates the deployment of applications created by third parties into a secure environment.

We believe the DCH platform vindicates the recent efforts of others to create open schemas for built environment sector semantic models. Our team is closely watching the evolution of standards such as Brick, RealEstateCore and ASHRAE 223P, with the ambition of offering a multi-schema building domain platform. The DCH makes strides with many challenges, but we recognise that using semantic models for re-deployable applications is not yet a solved problem. We hope that, by collaborating with the international community pursuing standardisation of built environment metadata schemas, the DCH platform can serve energy reduction projects of global significance.

## ACKNOWLEDGEMENTS

This work was funded by The Australian Renewable Energy Agency (ARENA), and by the Australian Institute of Refrigeration Air Conditioning and Heating (AIRAH), through the Affordable Heating and Cooling Innovation Hub (iHub).

## BIBLIOGRPAHY